\documentclass[useAMS,usenatbib]{mn2e}

\usepackage[dvips]{graphicx}

\usepackage[dvipsnames]{xcolor}
\usepackage{times,ulem}
\usepackage{multicol}
\usepackage{pdflscape}
\usepackage{slashbox}
\usepackage{soul}

\definecolor{naranja}{rgb}{1,0.5,0}

\definecolor{SeaBlue}{rgb}{0,0.7,0.3}

\definecolor{BlueSea}{rgb}{0,0.3,0.7}

\definecolor{purpu}{rgb}{0.65,0.1,0.5}

\defcitealias{2017MNRAS.467..340Z}{Paper\,I}

\title[Statistical tests in variability of AGNs II]{Microvariability
  in AGNs: study of different statistical methods \\II. Light curves from simulated images}
\author[L. Zibecchi et al.]{L.~Zibecchi$^{1,2,\star}$
  I.~Andruchow$^{1,2}$ S. A.~Cellone$^{1,3}$ and D. D.~Carpintero$^{1,2}$\\
$^{1}$ Facultad de Ciencias Astron\'omicas y Geof\'isicas, Universidad Nacional 
de La Plata, Paseo del Bosque, B1900FWA La Plata, Argentina.\\ 
$^{2}$ Instituto de Astrof\'isica de La Plata (IALP), CONICET-UNLP, 
Argentina.\\ 
$^{3}$ Complejo Astron\'omico El Leoncito (CASLEO), CONICET-UNLP-UNC-UNSJ, 
Argentina.\\
$^{\star}$ {\it Contact e-mail: lzibecchi@fcaglp.unlp.edu.ar}}

\begin{document}

\date{Accepted 2020 August 18. Received 2020 August 16; in original form 2020 April 22}

\pagerange{\pageref{firstpage}--\pageref{lastpage}} \pubyear{20XX}

\maketitle

\label{firstpage}

\begin{abstract}
In a previous paper, we studied two statistical methods used to analyse the variability of active galactic nuclei (AGNs): the $C$ and $F$ statistics. Applying them to observed differential light-curves of 39 AGNs, we found that, even though the $C$ criterion cannot be considered as an actual statistical test, it could still be a useful parameter to detect variability, whereas $F$ is a good detector of non-variability. In order to test these results under controlled input conditions, so that the different error sources could be individually evaluated, we generated a series of synthetic differential light curves simulating astronomical images with different atmospheric conditions, such as cloud cover, seeing or sky brightness, as well as several types of intrinsic variability of the AGN, all with a specific instrumental configuration. Having obtained light curves for each case, we applied both statistics to them in order to test their reliability. We found that a weight factor should always be used with these indices. The $F$ test has a tendency to classify noisy non-variable curves as variable (i.e. false positives), although it is reliable and robust to correctly classify non-variable curves. On the contrary, although the $C$ index tends to give false negatives, we found that whenever the $C$ index indicates a source to be variable, it effectively is. Finally, light curves with low amplitude variabilities are more likely to be affected by changes in atmospheric conditions.

\end{abstract}


\section{Introduction}
\label{sec:intro}

The phenomenon of variability in Active Galactic Nuclei (AGNs) is present throughout the entire electromagnetic spectrum, being especially prominent in blazars \citep[a sub-class of AGN with extremely collimated relativistic jets pointing within $\le $10$\degr$ to the line of sight, e.g.,][]{2014A&ARv..22...73F}. Its study provides important information about both the physical characteristics of the emission region and the parameters for different models. In particular, from the causality principle, the variability time scales ($\Delta t$) restrict the size of the emission region ($R$), through $R < c \Delta t$. The time scales involved range from months or years, i.e. long-term variability, going through days or weeks, i.e. short-term variability, to minutes or hours, i.e. microvariability or intra-night variability. The latter regime is mostly found in blazars, which can display variability time-scales (mainly in optical bands) in the order of minutes, that would imply an emission region smaller than the expected lower limit set by the SMBH event horizon. It is thus assumed that, in those cases, variability arises from enhanced emission from sufficiently small regions within the relativistic jet \citep[e.g.,][]{2008MNRAS.384L..19B}. In the optical range, variability is commonly detected through the statistical analysis of differential light curves (DLCs), which involve differential photometry between the AGN and suitably selected field stars (see  Sect. 2.1).

The detection of variations at scales of hours is affected by several effects: systematic errors introduced by contamination of the light from the host galaxy \citep{2000AJ....119.1534C}, inappropriate observational or photometric methodologies \citep{2007MNRAS.374..357C}, inadequate use of statistical methods for the detection \citep{2010AJ....139.1269D,2011MNRAS.412.2717J}, etc. Then, given the relevance of the phenomena, it is crucial to have reliable procedures and to use suitable statistical tests.

There are several works in the literature dedicated to the study and application of different tools for the detection of microvariability in light curves, such as: the $\chi^2$ test, which compares the distribution of the data in the light curve with the theoretical distribution of a non-variable object, which was proposed by \citet*{1976AJ.....81..919K}, and used for photometric and polarimetric time series \citep{2003A&A...409..857A,2005A&A...442...97A,2010AJ....139.1269D}; the One Way ANOVA test (Analysis of Variances), which consists on a family of tests that compare the means of a number of samples \citep{1998ApJ...501...69D,2004A&A...421...83R,2009AJ....138..991R,2010AJ....139.1269D,2017ApJ...849..161F,2017ApJS..232....7F,2019ApJ...880..155L,2019ApJ...871..192P}; the $C$ criterion, which involves the ratio of the standard deviations of two distributions \citep{1999A&AS..135..477R,2002A&A...390..431R,2005A&A...442...97A,2010AJ....139.1269D,2011MNRAS.412.2717J}; and the $F$ test, which takes into account the ratio between the variances of two distributions \citep{1988AJ.....95..247H,2010AJ....139.1269D,2011MNRAS.412.2717J}.

These tests simply compare the scatters of the target vs. comparison star, and control star vs. comparison star DLCs, thus relying on the assumption that measurement errors can be correctly represented by the scaled scatter of the latter. However, these methods disregard other valuable aspects already present in DLCs, like time-domain information \citep[e.g.,][]{S-1996}, which should be considered if one wants to construct a more sensible method for characterising AGN microvariability. Scatter methods are very popular in AGN variability studies because of their simplicity; however, the results are often contradictory when different tests are applied. Thus, it is important to firmly assess the reliability of the different statistical tools widely used by most authors (namely, the $C$ and $F$ statistics), finding which of them should be preferred under a wide range of conditions usually met in ground-based astronomical observations. In \citet[hereafter Paper I]{2017MNRAS.467..340Z}, we studied the $C$ and $F$ statistics with a large and homogeneous sample of real observational data, consisting of 78 nightly DLCs from 39 southern AGNs observed with the 2.15-m \lq Jorge Sahade\rq~telescope ({\sc CASLEO}, San Juan, Argentina). We found that, for DLCs with amplitudes $\Delta m$ near the rms error, the $F$ test is more prone than the $C$ criterion to classify them as variable, while for DLCs with larger amplitudes, both statistics tend to detect variability. With respect to the elapsed time $\Delta t$ corresponding to $\Delta m$, DLCs with large values of this parameter are more frequently classified as variable. Both statistics seem to be robust in the detection (or non-detection) of variability when the DLCs present low instrumental dispersion. We found that, even though the $C$ criterion cannot be considered as a theoretically well-grounded statistical test \citepalias[see][and references therein for details]{2017MNRAS.467..340Z}, it could still be a useful parameter to detect variability, provided that the correct significance factor is chosen. Thus, the $C$ criterion allows reliable variability results to be obtained, especially for small amplitude and/or noisy DLCs.

To analyse the reliability of different statistical tools usually employed in AGN variability studies, previous works relied on synthetic DLCs, where several known observational and atmospheric effects were included \citep{2010AJ....139.1269D,dD2014,2015AJ....150...44D,2013MNRAS.433..907E,2014Ap&SS.352...51W}. In those works, photon shot-noise and a Gaussian distribution of errors were assumed, though the latter is not the case for real, ground-based observations where the atmospheric and instrumental effects produce correlated errors with non-Gaussian distributions\footnote{In fact, error distributions in magnitude space ---where tests are usually applied--- are always non-Gaussian, even assuming Gaussian error distributions in flux.}. A different approach was developed by \citet{2000AJ....119.1534C}, who generated artificial images, taking into account different seeing conditions, in order to carry out a study on how seeing changes could lead to spurious variations in the differential magnitudes, when the flux contribution from the AGN host galaxy is not negligible. The addition of artificial stars to an observed field is a usual practice to assess completeness and photometric errors (see for example, \citet{Lee03} for an application to transits of extra solar planets). This approach can be similarly extended to galaxy images (e.g. \citet{Huang18}, where the authors studied how do variable seeing conditions affect photometric results).

In the present work, we obtained the light curves from artificial astronomical images generated to contemplate as real situations as possible. In those images, the typical observation features over an observing run were included, not only variations in the seeing, but also the presence of veil or clouds, the effects of the sky brightness due to the Moon, as well as the instrumental configuration, etc. (see Section \ref{sec:simul-imple} for more details).

We organized the paper as follows: in Section~\ref{sec:simul} we describe the implementation of the simulations and the generation of the synthetic differential light curves, and Section~\ref{sec:results} is devoted to the results. The discussion is presented in Section~\ref{sec:disc} and the conclusions in Section~\ref{sec:conclu}.

\section{Simulations}
\label{sec:simul}

\subsection{Statistical tools}
\label{sec:simul-tools}

We analysed both the $C$ criterion and the $F$ test. The former is defined as the ratio of the standard deviations of the data series to be compared; the latter, instead, is defined as the ratio of the variances of those data series \citepalias[see][for more details]{2017MNRAS.467..340Z}:
\begin{equation}
C=\frac {\sigma_{1}}{\sigma_{2}},
\end{equation}
\begin{equation}
F= \frac{\sigma^{2}_{1}}{\sigma^{2}_{2}},
\end{equation}
where, in our work, $\sigma_{1}$ ($\sigma^{2}_{1}$) {\bf is} the dispersion (variance) of the \lq object-comparison\rq~DLC\, and $\sigma_{2}$ ($\sigma^{2}_{2}$) that of the \lq control-comparison\rq~light curve. These DLCs are those used in the differential photometry technique, developed by \citet{1986PASP...98..802H}, which involves the object under study, plus one star used as comparison and another used as control. \citet{1988AJ.....95..247H} advised to take as control star one whose magnitude is close to that of the source, while the comparison star should be slightly brighter (the Howell's criterion). Hereafter, we will refer to the \lq object-comparison\rq~DLC as the \lq AGN DLC\rq~and to the \lq control-comparison\rq~DLC as the \lq control DLC\rq. The differences in magnitude between these three objects are taken into account by using a scaling factor, $\Gamma$ \citep[see][Eq. 13]{1988AJ.....95..247H}, which involves the target, comparison and control stars fluxes, the corresponding sky level, read-out noise, exposure time and the aperture area. The parameters change to:

\begin{equation}
C=\frac {\sigma_{1}}{\Gamma \sigma_{2}},
\end{equation}

\begin{equation}
F= \frac{\sigma^{2}_{1}}{\Gamma^{2} \sigma^{2}_{2}}.
\end{equation}

Throughout this work, the critical value for which the null hypothesis (i.e. statistical equality of dispersions/variances) is rejected is established at a significance level $\alpha = 0.995$. In the case of the $C$ criterion, the critical value is fixed at $2.576$, that would correspond  to a normal distribution with mean 0 and dispersion 1 (with a 99.5\% confidence level) if $C$ were distributed as a Gaussian. With respect to the $F$ test, the critical value is constructed from the significance level $\alpha$ and from the degrees of freedom (number of points in the light curve minus 1) of both DLCs involved. For more details, please refer to Paper I.

\subsection{Implementation}
\label{sec:simul-imple}

\begin{figure}
\centering
\includegraphics[width=0.4\textwidth]{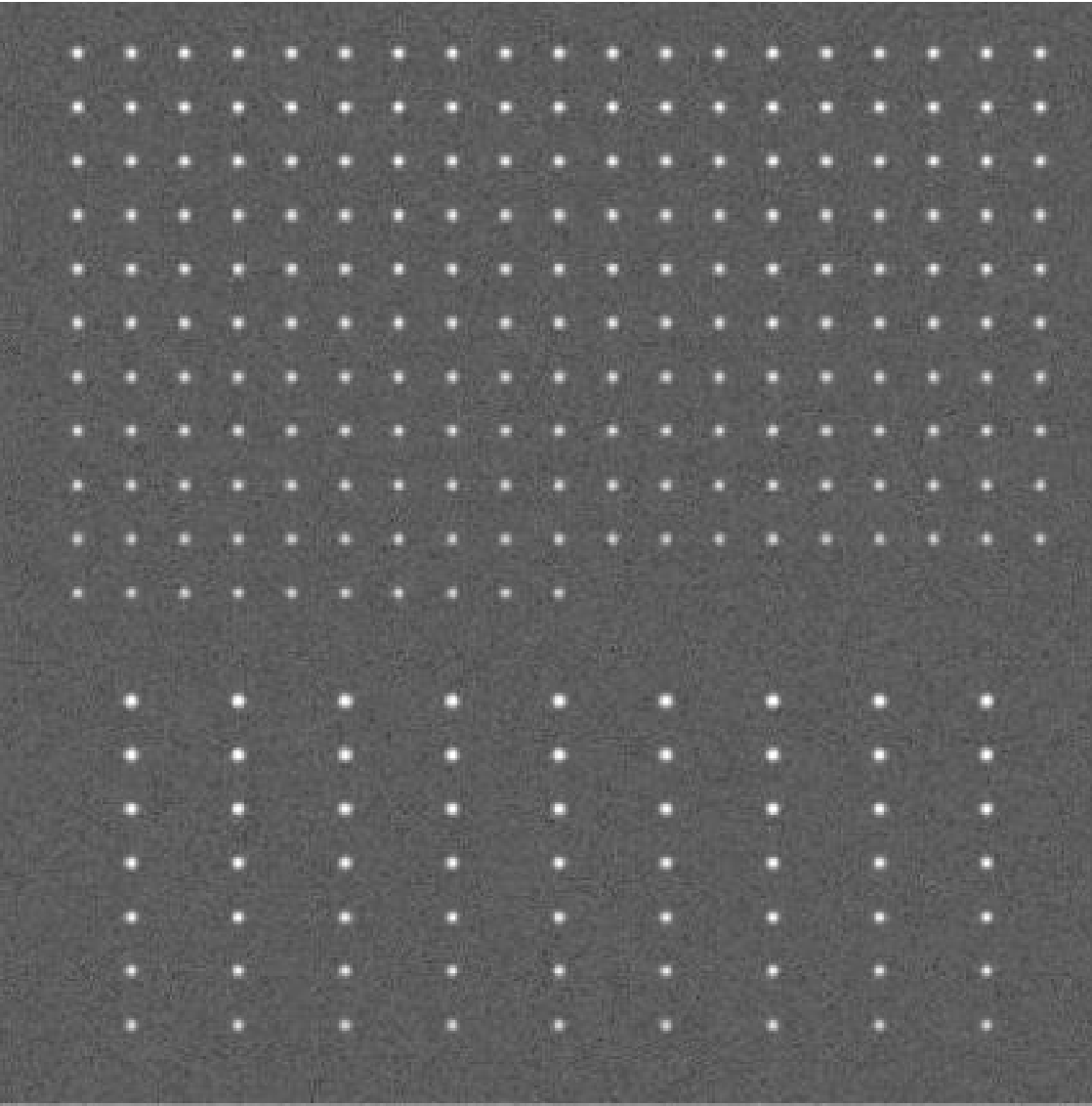}
\caption{Simulated CCD frame. Upper set: AGNs with different magnitudes. Lower set: candidates for comparison and control stars.}
\label{fig-simul}
\end{figure}

Based on several tasks of the software {\sc iraf}\footnote{{\sc iraf} is distributed by the National Optical Astronomy Observatories, which are operated by the Association of Universities for Research in Astronomy, Inc., under cooperative agreement with the National Science Foundation.} (Image Reduction and Analysis Facility), we developed a script that generates synthetic astronomical images. Since we required differential light curves, a set of point-like objects were placed in each frame. In Fig.~\ref{fig-simul} we show examples of these artificial objects. In the upper part there are 200 objects representing AGNs with magnitudes between 16 and 17 mag (in steps of 0.005 mag); this interval was chosen because it closely matches the low-magnitude regime of most AGN variability studies, including our own. An elliptical Moffat profile was chosen for the point spread function of the sources. In the lower part, there are 63 field stars, with magnitudes covering a range from 15 to 17 mag, used as comparison and control stars. The range of magnitudes corresponds to values close to those of standard stars in AGN fields \citep*{2001AJ....122.2055G}, so, this allowed us to apply Howell's criterion with several combinations among the AGN and the comparison and control stars. Standard magnitudes were converted into the corresponding counts (ADU) on the simulated CCD images considering the telescope and instrumental setup used to obtain the data analysed in \citetalias{2017MNRAS.467..340Z}. The Jorge Sahade telescope has a mirror size of 2.15 m, larger than most telescopes used in AGN variability studies; the reader should be aware of this when applying our results to evaluate variability studies which use telescopes of smaller diameters.

The script\footnote{The script is available upon request to the contact author.} included overheads (read-out time, etc.), which were used for the computation of the Universal Time assigned to each of the images, and also an additional component of random noise. We also included:

\begin{itemize}
 \item {\bf Instrumental noise:} related to the properties of the CCD. We adopted the values of the read-out noise and gain taken from the TeK1024 CCD at CASLEO, to match the sample in \citetalias{2017MNRAS.467..340Z}. The scale factor of the optical system is 0.813 arcsec/px.
 \item {\bf Seeing (IQ {\it Image quality}):} associated with the local atmosphere.
 \item {\bf Cloud cover (CC):} this effect simulates the extinction in magnitude caused by clouds, taking also into account those cases in which a veil is present. We took as the extinction its average during the exposure time.
 \item {\bf Sky brightness (SB)}: this takes into account the presence of the Moon and other light sources, affecting both the sky level and its associated rms noise.
  \item {\bf Airmass:} related to the altitude of the source.
\end{itemize}

The output resulted in a set of images with different observational and atmospheric conditions. In Table \ref{tab_simul} we show the values considered for each of these parameters. For the case of the airmass, we used a total of 40 values spanning from 1.2 to 2.0, with a uniform step. Combining the different effects, a total of 5000 frames (each containing 200 AGNs plus 63 stars) were generated.

The simulated atmospheric effects should be, in principle, cancelled out by the differential photometry technique\footnote{We are supposing that atmospheric effects are homogeneous throughout the relatively small simulated CCD field ($\approx 9 \times 9$ arcmin$^2$).}. Their net effect would then just be an increase of photometric errors, hence leading to higher DLCs dispersions. Systematic errors could however arise for extreme drops in S/N due to a combination of these effects \citep[e.g.][]{2007MNRAS.374..357C}. Other systematic effects affecting real observations, such as crowding, host-galaxy light contamination, defects in the CCD, flat-fielding residuals, poorly corrected cosmic rays hits, PSF variations across the field, and variations in the seeing produced by the possible imperfect guiding of the telescope were not taken into account. Early microvariability studies, in turn, show that errors arising from colour mismatch between the AGN and stars used to build the DLCs, coupled with differential extinction and airmass change, should be negligibly small \citep[e.g.,][]{1992AJ....104...15C}. Since we are simulating CCD images taken at CASLEO, we checked with the extinction coefficients for that site published in  \citet{2016BAAA...58..190F}, obtaining that any systematic effect on the differential $V$ magnitude should be $< 0.01$\,mag for a colour difference $\Delta (B-V)=1.0$ between AGN and comparison star, and for our full simulated airmass range ($\Delta \, \sec(z)=0.8$). So, we have not considered this effect either. Our results should then be taken as a general guide, to be complemented with those from real observations \citepalias{2017MNRAS.467..340Z}.

\begin{table}
\centering
\caption{Values of the instrumental and atmospheric conditions used in
  the simulations.} 
\label{tab_simul}
\begin{tabular}{ccccc}
\noalign{\medskip} \hline \hline \noalign{\smallskip}
Readout noise & Gain & Seeing & CC & SB \\
$e^-$ & $e^-$/adu & $''$ & mag & mag arcsec$^{-2}$ \\
\noalign{\smallskip} \hline \noalign{\smallskip}
9.60  & 1.98 & 0.6 & 0.00 & 22.2 \\
      &      & 1.5 & 0.25 & 21.2 \\
      &      & 2.0 & 0.50 & 20.7 \\
      &      & 3.0 & 0.75 & 20.2 \\
      &	     & 4.0 & 1.00 & 19.7 \\
\noalign{\smallskip}\hline
\end{tabular}
\end{table}

On all these images, we performed the usual reduction process with the {\sc iraf} packages. The tasks of the {\sc apphot} package were used for the aperture photometry. We selected an aperture radius of 8 pixels (equivalent to 6.5 arcsec), which is the radius at which the photometric growth curve stabilizes for all the seeing conditions considered, and for consistency with our real observations \citepalias{2017MNRAS.467..340Z}. The resulting photometry files were the input for a new \textsc{iraf} script which built the DLCs taking into account the different observational and atmospheric situations and the different combinations of the magnitudes of the objects. In this way, by selecting the same AGN on all the frames, we built a set of non variable DLCs. On the other hand, we also constructed  variable DLCs through an appropriate selection of different AGN images on the different frames; these simulated variable AGN curves were built following nine distinct variability patterns (or types):

\begin{enumerate}
\item {\it linear trend:} curves that correspond to a continuous increment (or decrement) of the magnitude throughout the entire observation. The decreasing linear trend variability was given an amplitude of 0.2 mag, while the increasing one was given an amplitude of 0.3 mag.
\item {\it flickering:} related to a random variation of the magnitude\footnote{The term \textsl{flickering} refers here to a stochastic variation, as defined in the radio band \citep{WW1995}.}. In total, five amplitudes of flickering were considered: 0.3, 0.2, 0.15, 0.10 and 0.05 mag (which we identify as flickerings 1 to 5, respectively).  
\item {\it wide peak:} represents a gentle increase in the flux followed by a mild decay with an amplitude of 0.15 mag.
\item {\it shark teeth-like:} two low amplitude (0.15 mag) bursts in a short time scale.
\end{enumerate}

\begin{figure}
\centering
\includegraphics[width=0.5\textwidth]{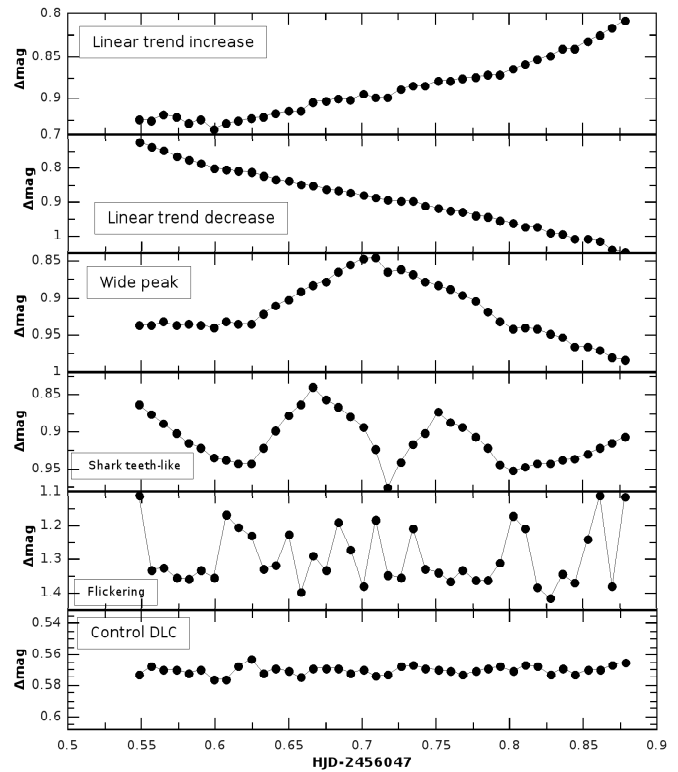}
\caption{Different types of variations, along with the control DLC. From top to bottom: increasing linear trend variability, decreasing linear trend variability, wide peak variability, shark teeth-like variability, flickering variability, and control DLC.}
\label{fig-curva}
\end{figure}

A first subset of variable DLCs ---including all nine patterns--- were generated from the synthetic images without atmospheric effects, except for the unavoidable airmass variation as the telescope tracks the target along the observation. These DLCs were considered as representative of the intrinsic variability behaviour of the AGNs (i.e., unaffected DLCs). Then, we used the images affected by simulated atmospheric effects in order to construct the DLCs that would allow us to study how the amplitude and shape of the intrinsic variations are modified under different atmospheric conditions (see Sect.~\ref{sec:simul-descr}). The different variability patterns are shown in Fig. \ref{fig-curva}, together with an example of a control DLC.

Given that, in some real observations, there are too few stars in the field, preventing the accurate application of Howell's criterion, we tested how a limited choice of comparison and control stars influences the results. To this end, the non-variable DLCs were subjected to three different restrictions on the allowed range of the magnitude differences between the AGN and the comparison star, and of the magnitude differences between both comparison and control stars, i.e. on the Howell's criterion. These restrictions were: difference in magnitudes between 0.001 and 0.1, between 0.1 and 0.3 , and between 0.3 and 0.5. For the variable DLCs, no restrictions were applied.

\subsection{Description}
\label{sec:simul-descr}

The conditions that commonly occur when observing AGNs with ground-based telescopes, were simulated considering exposure times of 120\,s plus the overheads, splitting the observation into 40 points at different airmasses, along a total simulated observed time of $\sim$8 h. In order to be able to compare the different situations, we included the effects one at a time, assuming different degrees of influence for each one. In this way, we built sets of DLCs as follows:

\begin{itemize}
\item{{\bf Control cases (CTR):} the Moon was absent (sky brightness 22.2 mag arcsec$^{-2}$, i.e. dark night), and no extinction by clouds was included. Five sets of DLCs, each for a fixed seeing along the entire night, were built (Table \ref{tab_simul}).}
\item{{\bf Variable seeing (IQ):} like the CTRs (five sets), but considering a variable seeing during the night.}
\item{{\bf Cloud cover (CC)}\footnote{The values of the extinctions were the average of the decrease in the magnitudes that occurred when the objects were observed with the presence of clouds.}: similar to the CTRs but including five different values of cloud cover. A total of 25 sets were obtained.}
\item{{\bf Sky brightness (SB):} similar to the CTRs, but with five different cases with the presence of the Moon, as seen in Table \ref{tab_simul}, and without clouds (25 sets).}
\item{{\bf Variable seeing and cloud cover (IQ-CC):} without the presence of the Moon. 25 sets were obtained from the combination of seeing and cloud variations.}
\item{{\bf Variable seeing and sky brightness (IQ-SB):} similar to the IQ-CC cases but with changes in the sky brightness and without clouds (25 sets).}
\item{{\bf Variable seeing, sky brightness and cloud cover (IQ-SB-CC):} 125 sets of DLCs built by taking into account the three effects together.}
\end{itemize}

\begin{figure}
\centering
  \includegraphics[width=0.45\textwidth]{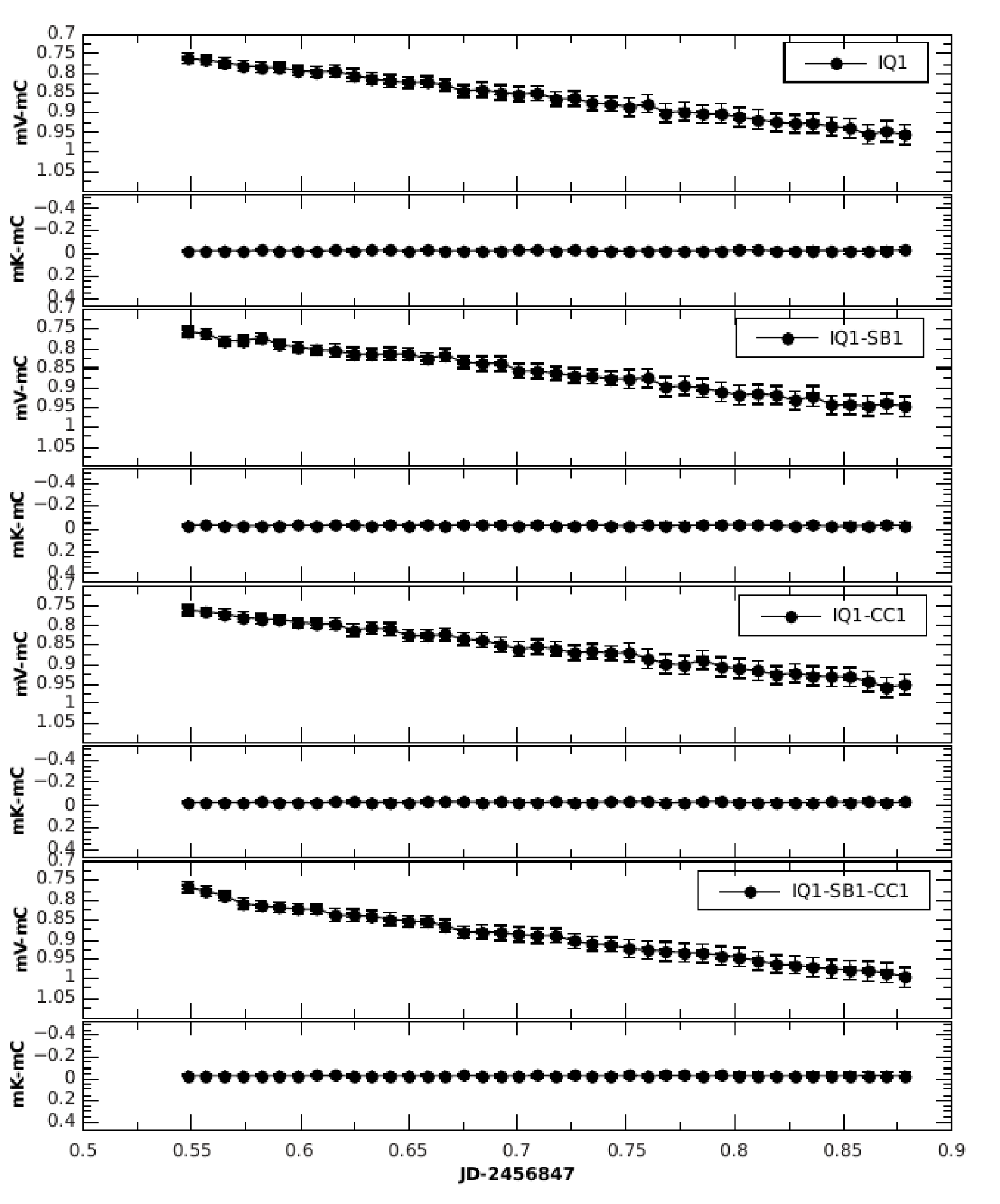}
  \caption{Examples of synthetic differential light curves for the situations IQ1, IQ1-CC1, IQ1-SB1 and IQ1-SB1-CC1. The mV-mC label in the y-axis corresponds to the AGN light curve, while the mK-mC label corresponds to the control curve. V is for the AGN, C for the comparison star and K for the control one, as defined in \citet{1986PASP...98..802H}.}
  \label{dlc-var1}
\end{figure}

A summary of the combinations of values taken for the image quality, cloud cover and sky brightness situations is shown in Table \ref{tab-situaciones-basicas}. The data points in each DLC were divided into five groups of eight points each, in order to apply the different combinations of the atmospheric effects. Taking into account the total of 235 sets of simulated situations, along with the three restrictions of the Howell's criterion for the non-variable AGNs, and the nine variability patterns (which were not subjected to the Howell's restrictions), and taking into account 200 possible AGNs and 63 field stars, a total of $5.6\times10^7$ synthetic DLCs were generated. As an example, we present in Fig. \ref{dlc-var1} a case of a variable AGN (decreasing linear trend variability) for four particular situations.

\begin{table*}
\centering
\caption{Details for the different situations of image quality, cloud cover and sky brightness  added to the simulated images. Column 1 shows the sets of eight points of the DLC. The different values of seeing, cloud cover and sky brightness are shown in columns 2 to 16. The last row indicates the mean value of the seeing for the IQs considered.}  
\label{tab-situaciones-basicas}
\begin{tabular}{|c|c|c|c|c|c|c|c|c|c|c|c|c|c|c|c|}
\noalign{\medskip} \hline \hline \noalign{\smallskip}
\small{\backslashbox{N}{Cases}} & \small{\bf IQ1} & \small{\bf IQ2} & \small{\bf 
IQ3} & \small{\bf IQ4} & \small{\bf IQ5} & \small{\bf CC1} & \small{\bf CC2} & 
\small{\bf CC3} & \small{\bf CC4} & \small{\bf CC5} & \small{\bf SB1} & 
\small{\bf SB2} & \small{\bf SB3} & \small{\bf SB4} & \small{\bf SB5} \\
\noalign{\smallskip} \hline \noalign{\smallskip}
 & \tiny{arcsec} & \tiny{arcsec} & \tiny{arcsec} & \tiny{arcsec} & \tiny{arcsec} 
& \tiny{mag} & \tiny{mag} & \tiny{mag} & \tiny{mag} & \tiny{mag} & 
\tiny{mag/arcsec$^2$} & \tiny{mag/arcsec$^2$} & \tiny{mag/arcsec$^2$} & 
\tiny{mag/arcsec$^2$} & \tiny{mag/arcsec$^2$} \\
\noalign{\smallskip} \hline \noalign{\smallskip}
\small{1-8}  & \small{1.5} & \small{1.5} & \small{2.0} & \small{3.0} & 
\small{0.6} & \small{0.0} & \small{0.0} & \small{0.25} & \small{0.75} & 
\small{0.0} & \small{19.7} & \small{20.5} & \small{22.2} & \small{21.1} & 
\small{22.2} \\
\noalign{\smallskip} \hline \noalign{\smallskip}
\small{9-16} & \small{0.6} & \small{2.0} & \small{2.0} & \small{3.0} & 
\small{0.6} & \small{0.25} & \small{1.0} & \small{0.25} & \small{0.50} & 
\small{0.0} & \small{19.7} & \small{21.1} & \small{21.1} & \small{22.2} & 
\small{21.1} \\
\noalign{\smallskip} \hline \noalign{\smallskip}
\small{17-24} & \small{1.5} & \small{2.0} & \small{3.0} & \small{2.0} & 
\small{1.5} & \small{0.50} & \small{0.0} & \small{0.50} & \small{0.25} & 
\small{0.25} & \small{19.7} & \small{22.2} & \small{20.5} & \small{22.2} & 
\small{21.1} \\
\noalign{\smallskip} \hline \noalign{\smallskip}
\small{25-32} & \small{2.0} & \small{3.0} & \small{4.0} & \small{2.0} & 
\small{1.5} & \small{0.75} & \small{0.75} & \small{0.25} & \small{0.0} & 
\small{0.50} & \small{19.7} & \small{22.2} & \small{20.5} & \small{22.2} & 
\small{21.1} \\
\noalign{\smallskip} \hline \noalign{\smallskip}
\small{33-40} & \small{2.0} & \small{3.0} & \small{3.0} & \small{3.0} & 
\small{2.0} & \small{0.75} & \small{0.0} & \small{0.25} & \small{0.0} & 
\small{0.0} & \small{19.7} & \small{22.2} & \small{20.5} & \small{22.2} & 
\small{21.1} \\
\noalign{\smallskip}\hline
\small{Mean value of IQ} & \small{1.5} & \small{2.3} & \small{2.8} & \small{2.6} 
& \small{1.2} & -- & -- & -- & -- & -- & -- & -- & -- & -- & -- \\
\noalign{\smallskip}\hline
\end{tabular}
\end{table*}

\section{Results}
\label{sec:results}

Although the goal of the synthetic DLCs was to explore some possible cases of observational and atmospheric parameters in a real night of observation, they do not represent an unbiased statistical sample of what one could expect when observing. Thus, the percentages reported here should not be taken as probabilities to be expected in a real campaign, but as an indication of the relative behaviour of the statistical tests under different observing conditions. The detailed analysis of all the results presented in this section, together with the corresponding figures and tables can be found in the appendix \ref{app-online} (on-line material).

\subsection{Influence of the scaling factor}
\label{sec:gamma}

\begin{figure}
\centering 
\includegraphics[width=0.46\textwidth]{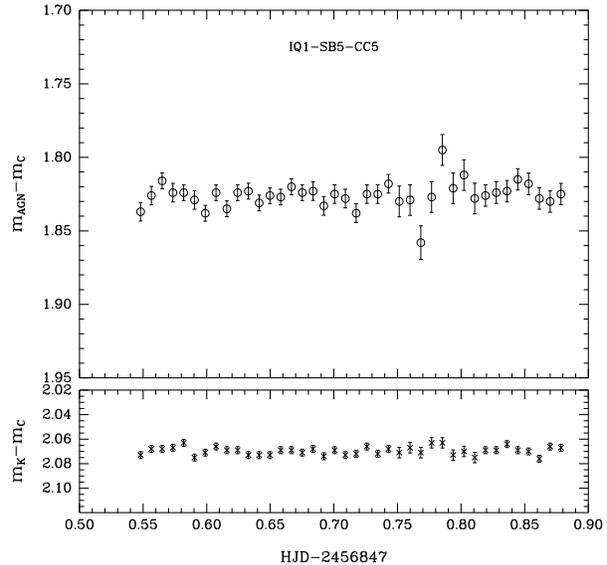}
\caption{Example of a differential light curve for a non-variable AGN (top) and the control DLC (bottom). The situation considered is IQ1-SB5-CC5. The mAGN-mC label in the $y$-axis corresponds to the AGN light curve, while the mK-mC label corresponds to the control curve.}
\label{curva-luz-agnnv}
\end{figure}

When working with observational data, the most common problem is the lack of stars in the field that meet Howell's criterion. Because of this, a value of the scaling factor $\Gamma$ was calculated for each set of parameters taken in the differential photometry. Therefore, $\Gamma$ was not unique for all the DLCs neither for all the objects \citepalias[see][]{2017MNRAS.467..340Z}. By definition, $\Gamma>1$ when the AGN is fainter than the control star, and $\Gamma<1$ when it is brighter (the role of the comparison star is not relevant to whether $\Gamma >1$ or $\Gamma<1$). We found this expected behaviour in our simulations. As an example of a choice of comparison and control stars not following Howell's criterion, we present in Fig. \ref{curva-luz-agnnv} a synthetic DLC where the use of $\Gamma$ is essential. We chose one of the non-variable AGNs with a value of instrumental magnitude of 17.255 and comparison and control stars with magnitudes of 15.429 and 15.398, respectively. The curve is one of the set which involves the first variable seeing, the fifth sky brightness and the fifth cloud cover situations (IQ1-SB5-CC5). When using the $C$ and $F$ parameters without the $\Gamma$ factor, both tools detected variability in the curve. When the scaling factor ($\Gamma=3.526$) was included, both parameters returned the correct non-variable state.

Using the control cases for the non-variable AGNs, we computed the percentages of variable and non-variable DLCs detected with $C$ and $F$, both when $\Gamma$ was included and when it was not. We found that the $C$ criterion recovered 100\% of the cases of non-variability of the AGNs, both with and without the scaling factor. Instead, for the $F$ test there were several hundreds of false positives without $\Gamma$, whereas this number dropped to tens when $\Gamma$ was applied, that is, the difference was over one order of magnitude. This behaviour also occurred when all the restrictions in the Howell's criterion were applied. Similar results were found when no restrictions were applied. All in all, the $F$ test resulted more sensitive than $C$ in (wrongly) classifying these non-variable DLCs as variable, especially when no $\Gamma$ weighting was applied. In view of these results (and in full agreement with \citetalias{2017MNRAS.467..340Z}), we strongly recommend using the factor $\Gamma$ for variability analysis. From now on, all the analysis will be made using this scaling $\Gamma$ factor.

\subsection{Number of points in the DLCs}
\label{sec:resul-n}

One of the main issues in variability studies is to obtain well-populated curves. This goal is generally not achieved for DLCs of astronomical sources, particularly AGNs, which are usually weak objects requiring relatively long exposures; moreover, in most ground-based studies any given source cannot always be followed throughout the entire night. The ideal situation would be to have hundreds to thousands of points in the curves \citep{2017MNRAS.464..274S}. However, in practice we usually have at most 40-50 points per curve (and substantially less in many microvariability studies). This makes the number of available variability indices limited, where the most appropriate tests to study light curves with a low number of points are those that involve the scatter of measurements \citep{2017MNRAS.464..274S}. In this sense, the $C$ criterion and the $F$ test are the most appropriate, especially for microvariability analysis.

We analysed how the variability results changed with the number of points in the DLCs. We studied DLCs with the original number of points considered in the simulations, $N=40$, and four additional cases with $N=30$, 20, 10 and 5 points, uniformly distributed along the observation night. We found that both, the $C$ criterion and the $F$ test, were stable in their variability classification down to $N=20$, both for variable and non-variable AGNs. An increment in the number of false positives and false negatives was observed in either indicator when using fewer points. In particular, the $C$ criterion was less stable and robust than the $F$ test. This is something to be expected because the $F$ test depends on the degrees of freedom ($N-1$), whereas the $C$ criterion does not (see Sect. 2.1).

The cases with fewer points in the DLCs gave the following results. For non-variable AGNs and $N=10$, the percentage of false positives was 0.3\% for both parameters. When $N=5$, this value reached 5\%. The dispersions increased for curves with decreasing number of points, especially in the AGN DLC. For the variable AGNs ---except the cases of flickering--- and $N=10$, both $C$ and $F$ always recovered 100\% of the variability state of the AGNs. With $N=5$, while the $C$ criterion still classified as variable 100\% of the DLCs, the $F$ test yielded 18\% of false negatives. Regarding the flickering variabilities, $N=10$ gave the same results as before. But for $N=5$, we found two main groups which behaved differently according to their original amplitude in magnitude. For the flickerings with $\Delta m>0.15$ mag, the $C$ parameter again classified 100\% of the DLCs as variable, whereas the $F$ test recovered almost 95\% of the variability state. On the other hand, for flickerings with $\Delta m<0.15$ mag, the numbers of false negatives were higher: with the $C$ criterion, the percentages of false negatives were between 0\% and 6\%, and for the $F$ test, between 11\% and 75\%. 

As it was expected, a well-sampled curve minimized the possibility of getting false results in the state of variability, while for light curves with small number of points, the performances of the $C$ and $F$ tests were poorer, which is likely due to a less accurate estimation of $\sigma$ and $\sigma^2$. Unless the amplitude of the variation is high, it is then not recommended to accept the statistical results without any additional consideration when working with less than 20 points.

\subsection{Behaviours with the IQ}
\label{sec:resul-iq}

We first analysed the relationship between the dispersions of the AGN and control DLCs and the scaling factor for the control and variable seeing cases of the non-variable AGNs. The comparison and control stars were not exactly the same in both cases, since seeing variations slightly change the total amount of light inside the apertures, thus leading the script to select a different star. Because of this, $\Gamma$ reached higher values in the cases of variable seeing, though this was not a problem since we took it into account in the statistical results of the $C$ and $F$ parameters. We found, for the three restrictions of Howell, that the $\sigma_1$ and $\sigma_2$ range values were the same for the control and variable cases. Also, there were no statistical differences in the distribution of values of $\sigma_1$ and $\sigma_2$ between the control and variable cases.

We repeated the foregoing computations for the case of variable AGNs, obtaining as before that the distributions of $\sigma_2$ were statistically indistinguishable, but ---obviously--- not so the distributions of $\sigma_1$. As expected, the range of values of $\sigma_2$ was the same in all cases since, for variable AGNs, the choice of the comparison and control stars was left free and all the possible combinations were considered. On the contrary, the set of values reached by $\sigma_1$ was different among the different cases because of their different original variability amplitudes and seeing conditions, the flickering 1 variability (that with the highest amplitude) being the most notable. Despite the differences in the distributions of $\sigma_1$, all the variable AGNs cases yielded 100\% of variable classifications with both $F$ and $C$, except that the $C$ criterion gave between 30\% and 70\% of false negatives for the flickering 5 (i.e., lowest amplitude) variability. 

Summarising, we found that it is the mean seeing value what really counts for the performance of the different tests, rather than seeing variations along the DLC.

\subsection{General results}
\label{sec:resul-gral}

We now study the behaviour of the DLCs when different combinations of changes in sky brightness and cloud cover, and seeing variations, are considered (Section \ref{sec:simul-descr} and Table \ref{tab-situaciones-basicas}). 

As an example, we present in the following the comparison between the second control case (CTR2, 1.50\,arcsec fixed seeing), the first variable seeing case (IQ1, 1.52\,arcsec mean seeing value), the first variable seeing plus the presence of the Moon case (IQ1-SB1), the first variable seeing plus first cloud cover changes case (IQ1-CC1), and the last case, including the first variation of seeing, clouds and the presence of the Moon (IQ1-SB1-CC1). Like in the previous sections, we started by studying the behaviour of the non-variable AGN DLCs. As the different effects were added, the magnitudes of the objects were affected, so, the range of values of $\Gamma$ increased. Notwithstanding this, the $\Gamma$ distribution is clearly around 1.00 for all the restrictions of Howell (see Section \ref{sec:simul-imple}). The values of $\sigma_2$ for the IQ1-SB1-CC1 cases were about three times higher than those from the CTR2 cases. Indeed, since $\sigma_2$ is a measure of the lack of quality of the observing run, it was expected to be higher when more effects are taken into account, making the DLCs noisier. The changes in sky brightness have a stronger effect on both, $\sigma_1$ and $\sigma_2$, than cloud cover variations; in each case, both dispersions seem to be equally affected. Adding the three effects together, $\sigma_1$ and $\sigma_2$ reached values that triple those that were obtained for the second control case. On the other hand, when different atmospheric situations were applied, the range of values of $\sigma_1$ and $\sigma_2$ increased and the range of the AGN magnitudes moved to weaker values. We found both changes whenever an atmospheric effect was added. These results give support to our confidence on the reliability of the simulations implemented in the present work.

The changes of $\sigma_1$, $\sigma_2$ and $\Gamma$ described in the previous paragraph were reflected in the results of the $C$ and $F$ tools. It is worth to notice that the $C$ parameter recovered 100\% of the DLCs as non-variable. In contrast, the $F$ parameter detected some DLCs as variable, although their percentage is low (less than 0.5\%). The highest percentage of these false detections corresponds to the restriction of Howell where the difference in magnitude is between 0.3 and 0.5. This was due to the fact that this restriction has the highest values of $\sigma_1$ and $\sigma_2$ and a significant number of AGNs weaker than the comparison stars, and all these factors contributed to make noisier DLCs. In this respect, we were obtaining again the results of \citetalias{2017MNRAS.467..340Z}, i.e. the $F$ test is more sensitive to the different observational error sources than the $C$ parameter.

For the case of the variable AGNs, $\sigma_2$ was affected in the same way as in the non-variable cases. On the other hand, the factor $\Gamma$ presented an increment in the high tail (from 2.60 to 3.10), which was largest when the effect of cloud cover was taken into account. Even so, the mean value of $\Gamma$ remained close to 1.00. The dispersion $\sigma_1$ also reached higher values when all the effects were included. The effects of the Moon presence have a larger incidence on $\sigma_2$, while $\sigma_1$ is more affected by the cloud cover. Considering all the effects together, the values of $\sigma_2$ spread along a range three times larger than that of the control case. It is clear that the control DLCs were relatively more affected by the atmospheric effects than the AGN DLCs, whose dispersions were already high due to the imposed variabilities.

For the foregoing cases of DLCs affected by the variations of the atmospheric conditions, both the $C$ and $F$ parameters recovered the original variability state of the AGNs. The cases of false negatives mostly occurred when all the atmospheric effects were involved. When the cloud cover effect was considered, the only variability affected was the flickering 5, whose original amplitude was 0.05 mag (comparable in order of magnitude to the expected noise in the control curves). In this case, both the $C$ and $F$ tools found non-variable DLCs, reaching the $C$ criterion a striking $\sim 80\%$ of the cases. On the other hand, the presence of the Moon is noticed not only in the flickering 5 case, but also in the shark teeth-like variability, as well as in the flickering 4, whose original amplitudes were 0.15 mag and 0.10 mag, respectively. For these types of variability, the $F$ test recovered 100\% of the state of variability for the DLCs, except for the flickering 5, while the $C$ criterion detected a small percentage of non-variable DLCs for the shark teeth-like and the flickering 4 variabilities, and a very high percentage for the flickering 5. Finally, when considering all the effects together, false negatives were detected whenever the original amplitude was $\Delta m<0.15$\,mag, irrespective of the variability type. In the case of the $F$ test, low values of false non-variable DLCs resulted for flickering 4, while for flickering 5 the percentage grew up to $\sim 70\%$. On the other hand, the $C$ parameter found non-variable DLCs for the five variabilities. As expected, a larger number of false non-variable cases were detected when the amplitude of the variability was smaller. This behaviour confirms that the $C$ parameter is too conservative when dealing with noisy DLCs. 

\subsubsection{Comparison between other cloud cover situations}
\label{sec:resul-II}

Taking the first cases of the variations in the seeing and sky brightness (IQ1-SB1) as a base, we analysed what happened when changes in the cloud cover occurred other than the first case CC1 (see Table \ref{tab-situaciones-basicas} for the types of cloud cover). For the non-variable AGNs, similar percentages were obtained in the statistics as in the CC1 cases. With each of the different restrictions, both the $C$ and $F$ parameters correctly classified 99.70\% of the DLCs. 

With respect to the variable AGNs, when considering only the cloud cover and the sky brightness variations it turned out that the only affected variability was the flickering 5, whose original amplitude was $\Delta m=0.05$ mag. In this last case, the difference was in the number of non-variable DLCs for each cloud cover. Going from the different cloud cover situations, the $F$ test recovered 99.6\% to 100\% of the variability. Applying the $C$ parameter, the percentage of variable DLCs was between 19\% and 60\%, the latter corresponding to the cloud cover situation, where a thin veil represented the mildest cloud cover and its mean dilution in the images was 0.15\,mag. This dilution was 0.45\,mag for the first cloud cover case, in which the false negatives were near 80\%. Since the cloud covers basically blocked up the light of the objects, making them weaker, this had a direct impact on the amount of noise of the DLCs. Thus, we found again that the $F$ test tended to classify noisy DLCs as variable, whereas the $C$ parameter did not detect the variability for curves with low amplitude, i.e. those most affected by changes in observational conditions. With the addition of the Moon, more of the variability types were affected in comparison to when each effect was taken individually. As we had noticed before, the most affected variabilities were those whose original amplitudes were $\Delta m < 0.15$ mag. 

\subsubsection{Comparison between image quality situations}
\label{sec:resul-III}

Finally, we analysed the results of the $C$ criterion and the $F$ test when adding to the cases of the previous section the variations in seeing other than the first case of variable seeing (Table \ref{tab-situaciones-basicas}). For the non-variable AGNs, the results of the $C$ and $F$ tools were similar to those already obtained: the $C$ criterion recovered 100\% of the variability state, while the $F$ test classified correctly 99.2\% of the DLCs. Again, for the variable AGNs, the most affected variabilities were those that had an original amplitude of $\Delta m < 0.15$ mag, specially for the $C$ parameter with the shark teeth-like, flickering 4 and 5 variabilities.

\subsubsection{Behaviour of the noisiest DLCs with the number of points on the curves}

As in Section \ref{sec:resul-n}, we analysed the statistical results when the number of points in the DLCs of the variable AGNs, $N$, changed from 40 to 30, 20, 10 and 5. As long as the variability amplitude of the AGNs was $\Delta m > 0.15$ mag and $N>20$, both the $C$ and $F$ tools recovered 100\% of the variability state for the DLCs of the variable AGNs. When $N$ dropped to 10, the false negatives of the $F$ test were 35\%,  and 20\% for the $C$ criterion. For $N=5$, the percentages were 25\% for the $C$ criterion and 75-80\% for the $F$ test. When the amplitude of the variability was less than 0.15 mag, false negatives were obtained even for DLCs with $N=40$, and the percentages became higher as the amplitude decreased. For $\Delta m = 0.15$ mag, these percentages for the $C$ criterion were less than 15\% for $N=40$, reaching 95\% for $N=5$. With respect to the $F$ test, the false negatives appeared when $N$ was less than 20, scaling up to 99\% when $N=5$. In all the cases, the highest percentages of false negatives corresponded to those situations where cloud cover variations were the most extreme. For $\Delta m = 0.1$ mag and $\Delta m = 0.05$ mag, false negatives reached numbers up to almost 100\%. The explanation for this behaviour can be found in a combination of changes in the dispersion $\sigma_2$ of the control DLC, which is affected by the different atmospheric effects, along with the lack of enough points in it, resulting in a higher value of $\sigma_2$. So, variable DLCs with few points and with large errors were statistically indistinguishable from the non-variable curves. On the other hand, we obtained the amplitude of the DLCs after being affected for the atmospheric conditions, $\Delta m_p$. When we analyzed the cases for a given pattern of variability affected with the same atmospheric conditions, the value of the amplitude $\Delta m_p$ was close to the originally proposed ($\Delta m$). We found this behaviour with the DLCs with $N > 10$. As the number of points was lower, the value of $\Delta m_p$ decreased with respect to the corresponding $\Delta m$. This happened for all the variability patterns. 

\section{Discussion}
\label{sec:disc}

By means of simulated CCD images including photometric error sources as well as different atmospheric effects (variable cloud cover, seeing, sky brightness), we studied the influence of several effects on the AGN DLCs and their variability state. The difference between the magnitude of the AGNs and that of the comparison and control stars used for differential photometry may lead to false results. This can be avoided by using the scaling factor $\Gamma$, as seen in Section \ref{sec:gamma}. \citet{2011MNRAS.412.2717J} proposed another weight factor, $\kappa$, which involves the ratio of the noise in the AGN DLC and the control DLC through their mean squared error. Like $\Gamma$, the factor $\kappa$ was defined to deal with the fact that the choice of the comparison stars may not be the ideal one. However, unlike $\Gamma$, $\kappa$ remains fixed for each object regardless its DLC. In this respect, the $\Gamma$ factor is more specific since it depends on each observing run and its calculation is based on the number of photons in the individual images \citep{1988AJ.....95..247H}. Therefore, it is more sensitive to changes in the observational conditions than $\kappa$. And, as we found, those conditions are the most important when analysing the state of variability of the source. The importance of applying a scaling factor to the statistical tools was already established in \citetalias{2017MNRAS.467..340Z} and is confirmed in the present work \citep[Section \ref{sec:gamma}; see also][]{2007MNRAS.374..357C}. Moreover, the results from our simulations are in good agreement with those obtained in \citetalias{2017MNRAS.467..340Z}, from the analysis of real observations.

Previous evaluations of statistical tests used to detect AGN variability \citep{2010AJ....139.1269D,dD2014,2015AJ....150...44D}, have generally concluded that $C$ is not a proper statistical test (about this particular, there was an extensive discussion in \citetalias{2017MNRAS.467..340Z}), while the $F$ and the ANOVA tests would be among the more suitable tools to correctly analyse the variability state of AGN DLCs. The differences between our work and previous studies based on simulated DLCs are the way in which they were built, the error sources considered, and their treatment. In particular, we used a wide variety of combinations among AGNs and stars for the differential photometry, making it possible to study, for example, the importance of using a scaling factor. Moreover, the way we built the DLCs allowed us to assume different types of intrinsic variabilities for the AGNs, with a variety of amplitudes, and including different atmospheric situations, photometric errors, etc. A number of new factors, not considered before, actually affecting the state of variability could also be studied, like the threshold in the original variability amplitude above which there were no errors in the variability classification of the DLCs. 

\citet{2017MNRAS.464..274S} found that the results obtained by using tests that involve dispersions are more reliable as the number of points increases. Our results do not only agree with this, but we also found that, if the number of points is less than 10, the statistical reliability of the results decreased no matter the state of variability or the amplitude of the original variability.

In \citetalias{2017MNRAS.467..340Z}, we studied DLCs built from photometry of field stars. Since we have all the night logs with the observing conditions for all the observations, we can compare those observations with the results of the non-variable AGNs of this work. We found that the number of false positives increased when the number of points decreased and when the night conditions were worse. These results are in total agreement with those in the present work. Also, the number of false positives (type-I errors) was always larger with the $F$ test than with the $C$ parameter, thus confirming that the former tends to classify noisy curves as variable. Conversely, type-II errors (false negatives) for noisy DLCs are common with the $C$ criterion, while quite infrequent with $F$. This contrasts with previous claims \citep{2010AJ....139.1269D, dD2014} of the $F$ test having a low power.

The large amount and variety of simulated DLCs built in the present work, allowed us to try a detailed comparison of observations and simulations, by finding ---in some cases--- a simulated curve that was close to a real DLC. Particularly, for curves with $N>10$, it was possible to reproduce both variable and non-variable observational DLCs. Regarding the observed AGN DLCs, in the cases were the night was photometric (absence of the Moon, no clouds and low values of seeing), all the combinations of field stars yielded non-variable DLCs for both tests (e.g. PKS\,1101$-$232, PKS\,2320+114). According to the results of Section \ref{sec:resul-gral}, these variability results are reliable. Another interesting case is PKS\,1622$-$297, observed during two nights with $N=13$ and $N=22$, respectively. Using the $C$ parameter, the AGN DLC resulted non-variable with all the possible combinations of the field stars. The $F$ test, however, classified the DLC as variable for the second night. Although the second night had more points, the seeing was better during the first one. Thus, according to the results of the present work, the $C$ criterion results are more reliable than those of the $F$ test, interpreting the last behaviour as a false positive due to a combination of $N\sim20$ and a noisy DLC. 

We also studied the possibility of having had false positives/negatives in Table 2 of \citetalias{2017MNRAS.467..340Z}. To this end, we inspected whether the night conditions were sufficient to explain a possible change in the state of variability of the DLCs. Analysing the $\sigma_1$, $\sigma_2$ and $\Gamma$ sets, we found that it is possible to have had these changes but only when all the factors are taken into account (i.e. situations of variable seeing, cloud cover and sky brightness), plus low variability amplitudes. We also found that it is more probable to have had false negatives than false positives, the former being variable AGNs observed through bad atmospheric conditions that masked the variability.

As for the differences between the $F$ test and the $C$ parameter, we take as an example the case for PKS\,0208$-$512. It was observed along two consecutive nights, 03-04 Nov. 1999, with 40 and 39 points in the DLC, respectively. According to the night logs, both nights had similar atmospheric conditions, with seeing around 2.5 to 3\,arcsec and with the presence of some veil and scattered clouds. The values of $\Gamma$ were close to one: 0.973 and 0.934; and those of $\sigma_2$ were 0.005 and 0.003. Using field stars with published standard magnitudes, it was possible to have the standard magnitude curves for the AGN. The peak-to-peak amplitudes were $\Delta m = 0.136$\,mag (first night) and 0.023 mag (second night), with $\langle m_{\mathrm v}\rangle =15.857 \pm 0.004$ and $\langle m_{\mathrm v} \rangle =15.814 \pm 0.004$, respectively, which yielded $\sigma_1= 0.046$ (night 1) and $\sigma_1=0.006$ (night 2). The source resulted variable in the first night according to both $C$ and $F$ tools, and non-variable for the $C$ criterion and variable for the $F$ test during the second night. Taking into account the results of Section \ref{sec:resul-gral}, we conclude that the reliable classification was that obtained with the $C$ parameter. On the other hand, though we do not have observations where the $F$ test yielded non-variability and the $C$ criterion variability, from the simulations we found that this may happen in those cases where the seeing varied and the number of points in the DLCs was $N<10$. In these cases, up to 3.42\% of the non-variable AGN DLCs were missclassified by the $C$ criterion, whereas up to 38.17\% were missclassified by the $F$ test. 

Finally, an important result was the one obtained with the shark teeth-like variability. This variability was built considering groups of eight points for each increasing/decreasing behaviour; the same eight point segments were considered for the variations of the different atmospheric parameters (Table \ref{tab-situaciones-basicas}). Even if such a situation where the intrinsic variations of the source and the atmospheric conditions are correlated/anti-correlated might be quite infrequent, it is valid to explore whether this could influence the DLCs classification. From the results obtained, we see that there were false negatives when applying the $C$ index. We found that in those cases the source was increasing its brightness but the cloud cover also increased, thus, while the dispersion of the control DLC increased, the dispersion of the AGN DLC decreased. 

\section{Conclusions}
\label{sec:conclu}

The variability state of a light curve allows us to know the physical mechanisms occurring at the source and to explain the observed behaviour. One drawback in the ground-based study of AGN microvariability is that their DLCs are usually not well sampled, especially for faint objects, which require  relatively long exposure times to achieve a sufficient S/N ratio with small to medium size telescopes. Our study is then based on simulated light-curves with a maximum number of 40 points, which is representative of most studies of intra-night variability. On the other hand, in order to obtain robust statistical results, tests need, in general, a significant amount of points, around one hundred or more. Therefore the statistical tools that can be used to study AGNs variability are limited. \citet{2017MNRAS.464..274S} pointed out that, when the number of points is less than 100, the statistical tests that involve dispersions are more stable (see their Fig. 5). Therefore, we chose the $C$ index and the $F$ test, both based on the dispersions of the DLCs, as tools to test on a series of simulated curves. The simulations were made by generating a series of images on which different instrumental and atmospheric conditions were included: changing airmass, seeing variations, different cloud coverage and different Moon phases. Based on these images, we built light curves where several errors present in real observations were included.

We found that a scaling factor should always be used. In particular, we analysed Howell's scaling factor \citep{1988AJ.....95..247H}, which takes into account the differences between the magnitudes of the objects involved in the differential photometry, and which is computed using the photon counts of the images. The inclusion of this factor turns out to be indispensable, regardless of the difference in magnitude.

When using statistical tests that involve dispersions, one of the crucial issues to obtain statistically supported variability results is to have a well-populated curve. The number of points in the DLC is thus also a relevant factor. We found that DLCs with at least 20 points are necessary to get reliable results. On the other hand, less than 10 points may yield false positives or false negatives, even in the best of situations (fixed and low seeing values, dark night, no clouds).

An important result we have found was that neither variations nor large seeing values (more than 3.0\,arcsec) influenced the robustness of the statistical tool used to classify the DLCs. In all cases (control and variable seeing cases), 100\% of the variability was recovered. In other words, and within the seeing values proposed for the simulations, we found that there were no differences between considering the values of the seeing point-to-point and taking an average value representative of it. Since our study was based on differential photometry and thus the seeing affected equally the source and all the field stars, it impacted on the quality of each individual image, not on the state of variability over an observing run. We note that the effect of crowding in the images, which may affect the foregoing conclusion, was not considered in our study because, in general, for the AGN fields we have the inverse problem (poorly populated star fields). We have also not considered the possible effects of the AGN host galaxies flux under variable seeing conditions since we are supposing that the host galaxy flux is non variable \citep[see][for a treatment of this effect]{2007MNRAS.374..357C}. Results may also be different if a smaller photometric aperture ---i.e., more sensitive to seeing changes--- is used.

When both the $C$ criterion and the $F$ test classify the DLCs as non-variable, we can trust in the obtained result. On the other hand, we also found that neither tool could distinguish real (but low-amplitude) variations from spurious ones due to atmospheric conditions introducing errors in the DLCs. This is because these external factors increase the dispersion of the DLCs, masking the AGNs intrinsic variability. In their study on the efficiency of different variability indices to detect variable stars, \citet{2017MNRAS.464..274S} found similar results, although they did not use either the $C$ index or the $F$ test, but a set of indices commonly used for the detection of variable stars.

Considering the different combinations of the variable seeing with the sky brightness, we found that the largest influence of the Moon occurs when there is a low-amplitude flickering (less than 0.10 mag), with the $C$ index yielding a larger number of false non-variable cases than the $F$ test. The same behaviour of the $C$ parameter was found when the cloud cover was added. The affected amplitudes were those lower than 0.05 mag; in this range and with the highest value of the seeing, the $C$ criterion yielded 90\% of false negatives. This is due to the combination of a change in the quality of the image with a decrease in the number of counts (presence of clouds), or an increase in the sky noise (presence of the Moon). This combination increases the noise in the DLCs masking out low-amplitude variations. This is reflected mostly on the $C$ index since, as seen in the non-variable AGNs, it never detects noise as variability.

Summarising, the $F$ test can yield false positives, but it is a good tool to detect non-variability. Vice versa, we found cases in which the $C$ index showed false negatives, though we can safely claim that a source is variable if the $C$ parameter indicates so. When it comes to detecting intra-night variability, the combination of the three atmospheric effects simultaneously in low amplitude variations could lead to masking the intrinsic variability present in the source, while each effect separately has a lower impact. In particular, when crowding and host galaxy light contribution are not issues, seeing changes have little effect on variability results. While these scatter methods give a simple means to obtain (with the limitations and caveats discussed here and in Paper\,I) reliable results, the next natural step will be to study specific methods to evaluate variability in AGN, taking into account time-domain information present in the light curves.

\section*{Acknowledgements}

The present work was supported by grant 11/G153 from the Universidad Nacional de La Plata. This research has made use of NASA's Astrophysics Data System. We want to thank E. Marchesini for the valuable help in this work, and the anonymous referee for the useful and positive suggestions to improve the present article.

\section*{Data availability}
Data available on request. The data underlying this article will be shared on reasonable request to the corresponding author.

\bibliographystyle{mn2e}
\bibliography{biblio}

\appendix
\section{On-line material}
\label{app-online}

\label{lastpage}

\end{document}